\documentclass[a4paper,10pt]{article}
\usepackage[utf8x]{inputenc}

\title{Deformation of the ABJM Theory}
\author{ Mir Faizal \\
 \\ 
Mathematical Institute, University of Oxford
\\ Oxford
OX1 3LB, United Kingdom 
 }

\begin{document}

\maketitle

\begin{abstract}
In this paper we analyse the ABJM theory on deformed spacetime. 
We show that this theory reduces to a
deformed  super-Yang-Mills theory when one of the scalar superfields 
is given a non-vanishing vacuum expectation value. Our analyse is done in 
$N=1$ superspace formalism.
\end{abstract}
\section{Introduction}
The low energy action for multiple $M2$-branes is thought to be given by the 
ABJM theory \cite{1, 2}.   This theory  is a
 three dimensional Chern-Simons-matter theory with gauge group
 $U(N)_k \times U(N)_{-k}$ at levels $k$ and $-k$ on 
the world-volume of $N$ $M2$-membranes placed at the fixed point of 
$R^8/Z_k$. This theory   explicitly realizes  $N=6$
 supersymmetry and it  is expected to be  
enhanced to full $N =8$ supersymmetry for $k =1,2$ \cite{su}. 
The ABJM theory coincides with the BLG theory
 for the only known example of 
 the Lie $3$-algebra \cite{BL1, BL2, BL3, blG, 5}. Both the BLG theory and the ABJM theory have been analysed in the 
 $N = 1$ superfield formalism  \cite{14,k1} and the Higgs mechanism for
both these theories has also 
 been studied in the $N =1$ superspace formalism \cite{14abjm, 14m}.

The spacetime noncommutativity arises in the string theory because of 
coupling the theory to a $NS$  background
 \cite{sw, dn}.
Other backgrounds  fields can similarly cause other deformations 
of the theory.  A non-anticommutative deformation of the 
theory is caused by a
$RR$ background  \cite{ov,se}.  
Furthermore, the commutator of the spacetime coordinates
with the fermionic coordinates 
does not vanish in the presence of  a gravitino background  
 \cite{bgn}.
Field theories with these kind of deformations 
 have  been 
thoroughly studied \cite{se}-\cite{bs}. 
In fact, it is well known that non-anticommutative deformation of a 
theory  breaks half of the supersymmetry of that theory explicitly. 
In this paper we will study those deformations of the $M$-theory which do not break any supersymmetry.

As there is a  duality between $M$-theory and $II$ string theory, 
we expect that any
 deformation on the $M$-theory side will also
correspond to some deformation on the string theory side.  
In fact,  $M2$-branes in $M$-theory  are analogous objects 
to strings in string theory. This is because just like 
strings can end on   $D$-brane in string theory, 
$M2$-branes can  end on $M5$-branes in $M$-theory. 
 Furthermore, as  a three-form field strength occurs
 naturally in $M$-theory, we expect that coupling 
the ABJM theory to a 
background three-form field could lead to a  
deformation of the ABJM theory, 
just like coupling of  $D$-branes to a background two-form  field 
strength leads to  a  noncommutative deformation of the theory.  
 
Coupling of the ABJM theory to background fields can have interesting 
uses in understanding the theory of multiple $M5$-branes. 
It may be noted that even though the action for a single
 $M5$-brane is known, the  
action   for multiple $M5$-branes is not
known \cite{41s}-\cite{42s}. 
Furthermore,  in $M$-theory the action of a single 
$M5$-brane can be obtained by analyzing the
the $\kappa$-symmetry of the open membrane ending on it \cite{14om}.
It might be possible to perform a similar analysis using 
the ABJM theory coupled to a background three-form field strength
and gain insight into the dynamics of multiple $M5$-branes.  

\section{ ABJM Theory in $N =1$ Superspace }
Now we review the 
 classical Lagrangian density  for the ABJM theory 
in $N =1$ superspace formalism
with the 
gauge group $U(N) \times U(N)$,
\begin{equation}
{ \mathcal{L}_c} =  \mathcal{L}_{M}
 + \mathcal{L}_{CS}
 - \tilde{\mathcal{L}}_{CS},
\end{equation}
where $\mathcal{L}_{CS}$ and 
 $\tilde{\mathcal{L}}_{CS}$ are deformed 
 Chern-Simons theories with gauge group's $U(N)$ from $U(N) \times U(N)$
respectively. They   can thus be expressed as  
\begin{eqnarray}
 \mathcal{L}_{CS} &=& \frac{k}{2\pi} \int d^2 \,  \theta \, \, 
  Tr \left[  \Gamma^a         \Omega_a
\right] _|, 
\nonumber \\
 \tilde{\mathcal{L}}_{CS} &=& \frac{k}{2\pi} \int d^2 \,  \theta \, \, 
  Tr \left[  \tilde{\Gamma}^a         \tilde{\Omega}_a
\right] _|, 
\end{eqnarray}
where $k$ is an integer and 
 \begin{eqnarray}
 \Omega_a & = & \omega_a - \frac{1}{6}[\Gamma^b, \Gamma_{ab}]    \\
 \omega_a & = & \frac{1}{2} D^b D_a \Gamma_b - \frac{i}{2} 
 [\Gamma^b , D_b \Gamma_a]    -
 \frac{1}{6} [ \Gamma^b ,
\{ \Gamma_b , \Gamma_a\}    ]   , \\
 \Gamma_{ab} & = & -\frac{i}{2} [ D_{(a}\Gamma_{b)} 
- i\{\Gamma_a, \Gamma_b\}    ],\nonumber \\
\tilde \Omega_a & = & \tilde \omega_a - \frac{1}{6}
[\tilde \Gamma^b, \tilde \Gamma_{ab}]    \\
 \tilde \omega_a & = & \frac{1}{2} D^b D_a \tilde \Gamma_b 
- \frac{i}{2}  [\tilde \Gamma^b , D_b \tilde\Gamma_a]    -
 \frac{1}{6} [ \tilde \Gamma^b ,
\{ \tilde \Gamma_b ,  \tilde \Gamma_a\} ]   ,  \\
 \tilde \Gamma_{ab} & = & -\frac{i}{2} [ D_{(a}\tilde \Gamma_{b)} 
- i\{\tilde \Gamma_a, \tilde \Gamma_b\}    ].
\end{eqnarray} 
Here the super-derivative $D_a$ is given by 
\begin{equation}
 D_a = \partial_a + (\gamma^\mu \partial_\mu)^b_a \theta_b,
\end{equation}
and  $'|'$  means that the quantity is evaluated at $\theta_a =0$. 
In component form the $\Gamma_a$ and $\tilde \Gamma_a$ are given by 
\begin{eqnarray}
 \Gamma_a = \chi_a + B \theta_a + \frac{1}{2}(\gamma^\mu)_a A_\mu + i\theta^2 \left[\lambda_a -
 \frac{1}{2}(\gamma^\mu \partial_\mu \chi)_a\right], \nonumber \\
 \tilde\Gamma_a = \tilde\chi_a + \tilde B \theta_a + \frac{1}{2}(\gamma^\mu)_a \tilde A_\mu + i\theta^2 \left[\tilde \lambda_a -
 \frac{1}{2}(\gamma^\mu \partial_\mu \tilde\chi)_a\right]. 
\end{eqnarray}
Thus in component form these Lagrangian are given by 
\begin{eqnarray}
  \mathcal{L}_{cs} &=& \frac{k}{4\pi}
\left( 2\left(\epsilon^{\mu \nu \rho} A_\mu     \partial_\nu  A_\rho 
 + \frac{2i}{3} A_\mu     A_\nu      A_\rho  \right)\right.\nonumber \\
  && \left. + 
 E^a     E_a + \mathcal{D}_\mu     ( \chi^a(\gamma^\mu)_a^b      E_b)\right),
\nonumber \\
   \tilde \mathcal{L}_{cs} &=& \frac{k}{4\pi}
\left( 2\left(\epsilon^{\mu \nu \rho}  \tilde A_\mu     \partial_\nu   \tilde 
A_\rho 
 + \frac{2i}{3}  \tilde A_\mu      \tilde A_\nu       \tilde A_\rho  
\right)\right. \nonumber \\ &&\left. + 
  \tilde E^a      \tilde E_a +  \tilde{\mathcal{D}}_\mu     
(  \tilde \chi^a(\gamma^\mu)_a^b       \tilde E_b)\right).
\end{eqnarray}

The Lagrangian density for the matter fields  is given by 
\begin{eqnarray}
 \mathcal{L}_{M} &=& \frac{1}{4} \int d^2 \,  \theta \, \,  
Tr \left[ [\nabla^a_{
}               X^{I \dagger}               
\nabla_{a 
}               X_I ] ] + 
  \mathcal{V}_{        } \right]_|,
\end{eqnarray}
where 
\begin{eqnarray}
 \nabla_{a}              X^{I } &=& D_a  X^{I } + i \Gamma_a          
    X^I - i  X^I        \tilde\Gamma_a      , \nonumber \\ 
 \nabla_{a}              X^{I \dagger} &=& D_a  X^{I  \dagger} 
- i X^{I  \dagger}       \Gamma_a    
        + i \tilde\Gamma_a            X^{I  \dagger}, 
\end{eqnarray}
and $\mathcal{V}      $ is the potential term  given by 
\begin{eqnarray}
 \mathcal{V}      & =& \frac{16\pi}{k}\epsilon^{IJ} \epsilon_{KL} 
[ X_I       X^{K \dagger}        X_J       X^{L\dagger}]. 
\end{eqnarray}

Now if the full finite  gauge 
transformation  under which this ABJM theory invariant  are given by
\begin{eqnarray}
   \Gamma_a \rightarrow i u {     } \nabla_a {     } u^{-1},&&
   \tilde \Gamma_a \rightarrow i \tilde u{     } 
 \nabla_a {     } \tilde u^{-1},\nonumber \\
 X^I \rightarrow  u  {     } X^I {     } \tilde u^{-1},&&
 X^{I\dagger} \rightarrow  \tilde u {     } X^{I\dagger} {     } u^{-1},
\end{eqnarray}
where 
\begin{eqnarray}
 u &=& [\exp ( i \Lambda^A T_A)]_{     }, \nonumber \\
\tilde u &=& [\exp ( i \tilde \Lambda^A T_A)]_{     }.
\end{eqnarray}
Thus now the infinitesimal gauge  transformations for these fields is given by  
\begin{eqnarray}
 \delta \Gamma_a =  \nabla_a {     } \Lambda, 
&&   \delta \tilde\Gamma_a = \nabla_a {     }
 \tilde\Lambda, \nonumber \\ 
\delta X^{I } = i(\Lambda{     } X^{I }  - X^{I }{     }\tilde \Lambda ),  
&&  \delta  X^{I \dagger  } 
= i(   \tilde \Lambda {     } X^{I\dagger  }-X^{I\dagger  }{     } \Lambda), 
\end{eqnarray}
The Lagrangian for the ABJM theory is invariant under these  gauge transformations 
\begin{eqnarray}
 \delta \mathcal{L}_{ABJM} &=&  \delta \mathcal{L}_{kcs} (\Gamma) -  
\delta \tilde{\mathcal{L}}_{-kcs} (\tilde\Gamma)  
+  \delta \mathcal{L}_M  \nonumber \\ &=& 0.
\end{eqnarray}

\section{Deformed   ABJM Theory }
In this section we shall construct a three dimensional 
  Chern-Simons   theory   
on a deformed superspace. 
In order to analyse deformation 
of the superspace both the Grassman 
coordinates and the spacetime coordinates are
 promoted to operators and a deformation of 
there superalgebra is imposed. In four dimensions
$N = 1$ supersymmetry is generated by the supersymmetric 
generator $Q_A$ which can be split into $Q_a$ and $Q_{\dot{a}}$. 
Furthermore, it is possible to break the supersymmetry corresponding 
to $Q_{\dot{a}}$ and leave the supersymmetry corresponding to $Q_a$
 intact or vice versa \cite{se}. Thus, it is possible to construct theories with 
 $N = 1/2$ supersymmetry in four dimensions. However, in three dimensions
 both $Q_a$ and $Q_{\dot{a}}$ act as independent supercharges.
 So, we can view a theory with $N =1$ supersymmetry in four 
dimensions as a theory with $N =2$ supersymmetry in three dimensions. 
Thus, $N = 1/2$ supersymmetry in four dimensions corresponds to $N =1$
 supersymmetry in three dimensions \cite{abjm1234}. It is not possible to obtain a 
$N = 1/ 2 $ theory in three dimensions as there are not enough degrees
 of freedom in three dimensions to do that. So, if we deform  a theory with
$ N =1$ supersymmetry in three dimensions, we can either 
retain all the supersymmetry or break all of it. However, we cannot
 partially break $ N =1$ supersymmetry in three dimensions. 
As supersymmetry is very important in the analysis of the ABJM 
theory we will deform the superspace  algebra in such a way 
that we do not break any  supersymmetry.
To do so we  promote ${ \theta}^a$ and ${y}^\mu$ 
 to operators $\hat{ \theta}^a$ and $\hat{y}^\mu$ 
which satisfy the following superspace algebra, 
\begin{eqnarray}
  [\hat{y}^\mu, 
\hat{y}^\nu] = 
 B^{\mu\nu}, &&
 {[\hat{y}^\mu, \hat{\theta}^a]}= A^{\mu a }.  
\end{eqnarray} 
We then use Weyl
ordering and  express the Fourier transformation of this superfield as, 
\begin{equation}
\hat{\Gamma}_a (\hat{y}, \hat{\theta}) =
\int d^4 k \int d^2 \pi e^{-i k \hat{y} -\pi \hat{\theta} } \;
\Gamma_a (k,\pi).
\end{equation}
Thus,  we  obtain  a one to one map between a function of
$\hat{\theta}, \hat{y}$ to a function of ordinary
 superspace coordinates $\theta, y$ via
\begin{equation}
\Gamma_a (y, \theta)  =
\int d^4 k \int d^2 \pi e^{-i k y -\pi \theta } \;
\Gamma_a (k,\pi).
\end{equation}
 We can express the product of two fields  
${\hat{\Gamma}^a}(\hat{y},\hat{\theta}) { \hat{\Gamma}_{a  }}
 (\hat{y},\hat{\theta})$
on this deformed superspace as
\begin{eqnarray}
{\hat{\Gamma}^a}(\hat{y},\hat{\theta}) { \hat{\Gamma}_{a  }}  
(\hat{y},\hat{\theta}) &=&
\int d^4 k_1 d^4 k_2 \int d^2 \pi_1  d^2 \pi_2
\exp -i( ( k_1 +k_2) \hat{y} +(\pi_1 +\pi_2) \hat{\theta} ) 
 \nonumber \\ && 
\,\,\,\,\, \,\,\,\,\,\, \,\,\,\, \,\,\,\,\,\, \,\,\,\, \,\,\,\,\,\, \,\,\,\, 
\,\,\,\,\,\, \times  \exp(i\Delta)
{\Gamma}^a  (k_1,\pi_1) {\Gamma}_{a  }   (k_2,\pi_2),
\end{eqnarray}
where
\begin{equation}
\exp (i\Delta) = \exp -\frac{i}{2} \left(
B^{\mu\nu} k^2_\mu k^1_\nu
+  A^{\mu a} (\pi^2 _a k^1_\mu - k^2_\mu \pi^1_a \right),
\end{equation}
This motivates the  definition of  the star product
  between ordinary functions, which is now defined as   
\begin{eqnarray}
{\Gamma^a}(y,\theta) \star { \Gamma_{a  }}  (y,\theta) & =& 
\exp -\frac{i}{2} \left(
 B^{\mu\nu} \partial^2_\mu \partial^1_\nu
+ A^{\mu a} (D^2 _a \partial^1_\mu -\partial^2_\mu 
D^1_a \right)) \nonumber \\ &&
\, \,\,\,\,\,\, \,\,\,\,\,\,\,\,\,\,
 \times 
 {\Gamma^a}(y_1,\theta_1) { \Gamma_{a  }}  (y_2, \theta_2)
\left. \right|_{y_1=y_2=y, \; \theta_1=\theta_2=\theta}.
\label{star2}
\end{eqnarray}
Here we have defined the star product between ordinary functions using 
super-derivative $D_a$ rather than $\partial_a$ because they commute 
with the generators of the supersymmetry $Q_a$ \cite{abjm1234},
\begin{equation}
 \{ Q_a, D_b \} =0.
\end{equation}
Thus, to write the ABJM in this deformed superspace, we now use
 \begin{eqnarray}
 \hat \Omega_a & = & \hat \omega_a - \frac{1}{6}[\hat \Gamma^b, \hat
\Gamma_{ab}]\nonumber \\ 
& = & \omega_a - \frac{1}{6}[\Gamma^b, \Gamma_{ab}]_\star  \\
 \hat \omega_a & = & \frac{1}{2} D^b D_a \hat \Gamma_b - \frac{i}{2} 
 [\hat \Gamma^b , D_b \hat\Gamma_a] -
 \frac{1}{6} [ \hat \Gamma^b ,
\{ \hat \Gamma_b , \hat \Gamma_a\} ]
\nonumber \\&=&  \frac{1}{2} D^b D_a \Gamma_b - \frac{i}{2} 
 [\Gamma^b , D_b \Gamma_a]_\star -
 \frac{1}{6} [ \Gamma^b ,
\{ \Gamma_b , \Gamma_a\}_\star ]_\star,  \\
 \hat\Gamma_{ab} & = &-\frac{i}{2} [ D_{(a}\hat \Gamma_{b)} 
- i\{\hat \Gamma_a, \hat \Gamma_b\}) \nonumber \\ &=& -\frac{i}{2} ( D_{(a}\Gamma_{b)} 
- i\{\Gamma_a, \Gamma_b\}_\star ], \\
\hat {\tilde \Omega}_a & = & \hat {\tilde \omega}_a - \frac{1}{6}
[\hat { \tilde \Gamma}^b, \hat {\tilde \Gamma}_{ab}] 
\nonumber \\ &=&
\tilde \omega_a - \frac{1}{6}
[\tilde \Gamma^b, \tilde \Gamma_{ab}]_\star   \\
 \hat {\tilde \omega}_a & = &\frac{1}{2} D^b D_a \hat { \tilde \Gamma}_b 
- \frac{i}{2}  [\hat { \tilde \Gamma}^b , D_b \hat {\tilde\Gamma}_a]
 -
 \frac{1}{6} [ \hat {\tilde \Gamma}^b\{ \hat {\tilde \Gamma_b} , 
 \hat {\tilde \Gamma}_a\} ]
 \nonumber \\ &=&  \frac{1}{2} D^b D_a \tilde \Gamma_b 
- \frac{i}{2}  [\tilde \Gamma^b , D_b \tilde\Gamma_a]_\star -
 \frac{1}{6} [ \tilde \Gamma^b ,
\{ \tilde \Gamma_b ,  \tilde \Gamma_a\}_\star ]_\star,  \\
 \tilde \Gamma_{ab} & = &-\frac{i}{2} [ D_{(a}\hat {\tilde \Gamma}_{b)} 
- i\{\hat {\tilde \Gamma}_a, \hat {\tilde \Gamma}_b\}]
\nonumber \\&=& 
 -\frac{i}{2} [ D_{(a}\tilde \Gamma_{b)} 
- i\{\tilde \Gamma_a, \tilde \Gamma_b\}_\star].
\end{eqnarray}  
In order to analyse the gauge transformations of this deformed 
ABJM theory it will be useful to define 
\begin{eqnarray}
 u &=&[\exp ( i \hat\Lambda^A T_A)] = [\exp ( i \Lambda^A T_A)]_{\star  }, \nonumber \\
\tilde u &=&[\exp ( i \hat{\tilde \Lambda}^A T_A)] = [\exp ( i \tilde \Lambda^A T_A)]_{\star  }.
\end{eqnarray}
The star product reduces to the usual Moyal star product
for the bosonic noncommutativity  in the limit $C^{ab}=A^{a\mu}=0$ and 
for $A^{a\mu}=C^{ab}=0$ it reduces to the standard 
fermionic star product. 

Now we construct the 
 classical Lagrangian density with the 
gauge group $U(N) \times U(N)$,
  on this deformed superspace, 
\begin{equation}
{ \mathcal{L}_c} =  \mathcal{L}_{M}
 + \mathcal{L}_{CS}
 - \tilde{\mathcal{L}}_{CS},
\end{equation}
where $\mathcal{L}_{CS}$ and 
 $\tilde{\mathcal{L}}_{CS}$ are deformed 
 Chern-Simons theories with gauge group's $U(N)$ from $U(N) \times U(N)$
respectively. They   can thus be expressed as  
\begin{eqnarray}
 \mathcal{L}_{CS} &=& \frac{k}{2\pi} \int d^2 \,  \theta \, \, 
  Tr \left[  \hat \Gamma^a     \hat\Omega_a
\right] _| \nonumber \\ &=& \frac{k}{2\pi} \int d^2 \,  \theta \, \, 
  Tr \left[   \Gamma^a  \star    \Omega_a
\right] _| , 
\nonumber \\
 \tilde{\mathcal{L}}_{CS} &=& \frac{k}{2\pi} \int d^2 \,  \theta \, \, 
  Tr \left[  \hat{\tilde{\Gamma}}^a     \hat{\tilde{\Omega}}_a
\right] _|\nonumber \\ &=& \frac{k}{2\pi} \int d^2 \,  \theta \, \, 
  Tr \left[  \tilde{\Gamma}^a  \star    \tilde{\Omega}_a
\right] _|.  
\end{eqnarray}
The Lagrangian  for the matter fields  is given by 
\begin{eqnarray}
 \mathcal{L}_{M} &=& \frac{1}{4} \int d^2 \,  \theta \, \,  
Tr \left[ [\nabla^a_{
}           \hat X^{I \dagger}           
\nabla_{a 
}           \hat X_I ] ] + 
  \hat \mathcal{V}_{    } \right]_|,
\end{eqnarray}
where 
\begin{eqnarray}
 \nabla_{a}    \hat      X^{I } &=& D_a  \hat X^{I } + 
i \hat \Gamma_a   \hat   
    X^I - i  \hat X^I \hat{   \tilde\Gamma}_a      , \nonumber \\ 
 \nabla_{a}   \hat        X^{I \dagger} &=& D_a \hat X^{I  \dagger} 
- i \hat X^{I  \dagger} \hat   \Gamma_a    
        + i \hat{\tilde\Gamma}_a   \hat      X^{I  \dagger}, 
\end{eqnarray}
and $\hat \mathcal{V}  $ is the potential term  given by 
\begin{eqnarray}
 \hat \mathcal{V}   & =& \frac{16\pi}{k}\epsilon^{IJ} \epsilon_{KL} 
[\hat X_I\hat X^{K \dagger} \hat    X_J \hat  X^{L\dagger}]. 
\end{eqnarray}
Now we can also express the Lagrangian for the matter fields as 
\begin{eqnarray}
 \mathcal{L}_{M} &=& \frac{1}{4} \int d^2 \,  \theta \, \,  
Tr \left[ [\nabla^a_{
}  \star          X^{I \dagger}     \star       
\nabla_{a 
}  \star          X_I ] ] + 
  \mathcal{V}_{\star     } \right]_|,
\end{eqnarray}
where 
\begin{eqnarray}
 \nabla_{a}    \star       X^{I } &=& D_a  X^{I } + i \Gamma_a   \star    
    X^I - i  X^I \star    \tilde\Gamma_a      , \nonumber \\ 
 \nabla_{a}   \star        X^{I \dagger} &=& D_a  X^{I  \dagger} 
- i X^{I  \dagger} \star   \Gamma_a    
        + i \tilde\Gamma_a   \star      X^{I  \dagger}, 
\end{eqnarray}
and $\mathcal{V}_\star   $ is the potential term  given by 
\begin{eqnarray}
 \mathcal{V}_\star   & =& \frac{16\pi}{k}\epsilon^{IJ} \epsilon_{KL} 
[ X_I\star    X^{K \dagger} \star    X_J \star   X^{L\dagger}]. 
\end{eqnarray}

Now if the full finite  gauge 
transformation  under which this ABJM theory invariant  are given by
\begin{eqnarray}
   \Gamma_a \rightarrow i u {\star  } \nabla_a {\star  } u^{-1},&&
   \tilde \Gamma_a \rightarrow i \tilde u{\star  } 
 \nabla_a {\star  } \tilde u^{-1},\nonumber \\
 X^I \rightarrow  u  {\star  } X^I {\star  } \tilde u^{-1},&&
 X^{I\dagger} \rightarrow  \tilde u {\star  } X^{I\dagger} {\star  }
 u^{-1}.
\end{eqnarray}
Thus now the infinitesimal gauge  transformations for these fields is given by  
\begin{eqnarray}
 \delta \Gamma_a =  \nabla_a {\star  } \Lambda, 
&&   \delta \tilde\Gamma_a = \nabla_a {\star  }
 \tilde\Lambda, \nonumber \\ 
\delta X^{I } = i(\Lambda{\star  } X^{I }  - X^{I }{\star  }\tilde \Lambda ),  
&&  \delta  X^{I \dagger  } 
= i(   \tilde \Lambda {\star  } X^{I\dagger  }-X^{I\dagger  }{\star  } \Lambda), 
\end{eqnarray}
The Lagrangian for the ABJM theory is invariant under these  gauge transformations 
\begin{eqnarray}
 \delta \mathcal{L}_{ABJM} &=&  \delta \mathcal{L}_{kcs} (\Gamma) -  
\delta \tilde{\mathcal{L}}_{-kcs} (\tilde\Gamma)  
+  \delta \mathcal{L}_M  \nonumber \\ &=& 0.
\end{eqnarray}
\section{Higgs Mechanism}
Now we take the vacuum expectation value of one of the scalar 
superfields say $X$, to be a non-zero,
\begin{equation}
<{X}> =\nu \neq 0. 
\end{equation}
This  spontaneously breaks the symmetry from 
 $U(N)\times U(N)$ to its diagonal subgroup, $U(N)$.  Now let
$A_a$ be superfield associated with the broken gauge and 
$B_a$ be associated with the unbroken gauge group. Then we have   
\begin{eqnarray}
A_{a}&=&\frac{1}{2}\left( \Gamma_a -\tilde \Gamma_a\right),\nonumber \\
B_{a}&=&\frac{1}{2}\left( \Gamma_a + \tilde \Gamma_a\right),
\end{eqnarray}
Now we can write the Chern-Simons part of the Lagrangian  as 
\begin{eqnarray}
  \mathcal{L}_{CS} &=& \frac{k}{2\pi} \int d^2 \,  \theta \, \, 
  Tr \left[  A^a  \star \left[  W_a  + \frac{1}{6}[A^b, A_{ab}]_\star
 \right]
\right] _|,
\end{eqnarray}
where 
\begin{eqnarray}
 W_a &=& \frac{1}{2} D^b D_a B_b - \frac{i}{2} 
 [B^b , D_b B_a]_\star -
 \frac{1}{6} [ B^b ,
\{ B_b , B_a\}_\star ]_\star,\nonumber \\
A_{ab} & = & -\frac{i}{2} [ D_{(a}A_{b)} 
- i\{A_a, A_b\}_\star ].
\end{eqnarray}
Now as the gauge group is broken down to its diagonal subgroup, we can 
integrate out the field $A_a$ by using its equations of motion. 
Using  this value of $A_a$ thus obtained we get the a theory
 whose first term corresponds to noncommutative  super-Yang-Mills theory. 
Thus the kinetic part of this theory has the terms 
\begin{equation}
 \mathcal{L}_{YM} = \frac{k^2}{16\pi^2 \nu^2}
 \int d^2 \,  \theta \, \,  [W^a \star W_a  + [\mathcal{O}]_\star]_|.
\end{equation}
There are higher derivative terms in this action corresponding to different
whose origin is not the noncommutative nature of the theory we have analysed.
Similarly in the kinetic part of the matter field also contains higher 
derivatives. However, the first term is a usual  gauge theory term  with 
the covariant derivatives corresponding to the unbroken gauge field $B_a$.
\begin{equation}
  \mathcal{L}_{KM} = \frac{k^2}{16\pi^2 \nu^2} 
 \int d^2 \,  \theta \, \, 
[\nabla^a_{
}  \star          X^{I \dagger}     \star       
\nabla_{a 
}  \star          X_I
 + [\mathcal{O}]_\star]_|,
\end{equation}
where 
\begin{eqnarray}
 \nabla_a\star X^I & =&  D_a X^I -i B_a \star X^I, \nonumber \\ 
\nabla_a\star X^{I\dagger} & =&  D_a X^{I\dagger} + i B_a \star X^{I\dagger}.
\end{eqnarray}
The potential term can be now rewritten  as $V_\star$ using 
 equation of motion of $A_a$. 
We identify the Yang-Mills coupling with 
\begin{equation}
 g_{YM} = \frac{2\pi \nu}{k}. 
\end{equation}
It is possible to keep
$g_{YM}$ fixed in   the limit $\nu \to \infty$ and $k \to \infty$. So we 
will now only consider the leading order terms in powers of $\nu$ and $k$.
 thus we can write the full Lagrangian as 
\begin{equation}
 \mathcal{L}_T = \frac{1}{g_{YM}^2}  
 \int d^2 \,  \theta \, \, [ W^a \star W_a 
+\nabla^a_{
}  \star          X^{I \dagger}     \star       
\nabla_{a 
}  \star          X_I + V_\star ]_|
\end{equation}

Now if the full finite  gauge 
transformation  under which this theory is invariant  are given by
\begin{eqnarray}
 B_a \rightarrow i v {\star  } \nabla_a {\star  } v^{-1},&&
 X^I \rightarrow  v  {\star  } X^I ,\nonumber \\ 
 X^{I\dagger} \rightarrow   X^{I\dagger} {\star  } v^{-1},&& 
\end{eqnarray}
where 
\begin{equation}
 v = u + \tilde u.
\end{equation}
We can write now $v$ as follows 
\begin{equation}
 v = [\exp (i \lambda^A T_A)]_\star.
\end{equation}
It may also be noted that 
\begin{equation}
 \delta W_a = v \star W_a \star v^{-1},
\end{equation}
Now the infinitesimal gauge  
transformations for these fields can be written as 
\begin{eqnarray}
 \delta B_a =  \nabla_a {\star  } \lambda,   && 
\delta X^{I } = i\lambda{\star  } X^{I }, \nonumber \\  
  \delta  X^{I \dagger  } 
= -i   X^{I\dagger  }{\star  }\lambda. 
\end{eqnarray}
The Lagrangian for the ABJM theory is invariant under these  gauge transformations 
\begin{equation}
 \delta \mathcal{L}_{T} = 0.
\end{equation}
Thus this theory is just a super-Yang-Mills theory deformed by the 
background fields. This deformed 
super-Yang-Mills theory arises a result of coupling  $D$-branes
 to background 
fields. This motivates the question if it is 
possible to write a non-linear Born-Infidel type extension to ABJM theory.
 As the gauge part is purely topological we expect no change in that part. 
However, it might be possible to write  the kinetic part of the matter 
fields as a Born-Infidel type theory.

\section{Conclusion}
In this paper we have analysed the ABJM theory in $N =1$ 
superspace
on noncommutative spacetime. It was show that this theory is invariant under 
noncommutative gauge transformations, which in commutative limit reduce to
regular gauge transformations for the ABJM theory. Furthermore, 
it was demonstrated that this theory reduces to the noncommutative
 super-Yang-Mills
theory when one of the scalar fields is given an vacuum expectation value. 

 Recently  supersymmetric
Chern-Simons theory  has also 
been used to study fractional quantum Hall effect via holography \cite{e}
and analyse the $AdS_4/CFT_3$ correspondence \cite{6}-\cite{10}.  
 It will be interesting  to
 analyse similar effects for 
this deformed ABJM theory. It will also be interesting to analyse the
 BRST and the anti-BRST symmetries 
of this model. These symmetries for noncommutative deformation of the 
ABJM theory in various have been already analysed \cite{k1}. It will be 
interesting to analyse the BRST and the anti-BRST symmetries of the 
ABJM theory with non-anticommutative deformation.


\begin{thebibliography}{99}
 \bibitem{1}I. L. Buchbinder,
 E.A. Ivanov, O. Lechtenfeld, N. G. Pletnev, I. B. Samsonov and B. 
M. Zupnik, 
	JHEP. 0903, 096 (2009)
 \bibitem{2}O. Aharony, O. Bergman, D. L. Jafferis and J. Maldacena, 
 JHEP. 0810, 091 (2008) 
\bibitem{su}
O-Kab Kwon, P. Oh and  J. Sohn, JHEP.  0908, 093 (2009)
\bibitem{BL1} J. Bagger and N. Lambert, Phys. Rev. D75, 045020 (2007)
\bibitem{BL2}J. Bagger and N. Lambert, Phys. Rev. D77, 065008 (2008)
\bibitem{BL3}J. Bagger and N. Lambert, JHEP. 0802, 105 (2008)
\bibitem{blG}A. Gustavsson, Nucl. Phys. B811, 66 (2009)
 \bibitem{5}M. A. Bandres, A. E. Lipstein and J. H. Schwarz, JHEP. 0809, 027 (2008)
\bibitem{14}A. Mauri and  A. C. Petkou, Phys. Lett. B666,  527 (2008)
\bibitem{k1}
M. Faizal,  Phys. Rev. D84, 106011 (2011) 
\bibitem{14abjm}
S. V. Ketov and S. Kobayashi, Phys. Rev. D83, 045003 (2011) 
\bibitem{14m}
S. Mukhi and  C. Papageorgakis, JHEP. 0805, 085 (2008)
 \bibitem{sw} N. Seiberg and E. Witten, 
 JHEP. 9909, 032 (1999)
\bibitem{dn}
M. R. Douglas and N. A. Nekrasov,  Rev.  Mod.  Phys. 73, 977 (2001)
\bibitem{dfr} S. Doplicher, K. Fredenhagen and  J. E. Roberts,
    Commun. Math. Phys. 172, 187 (1995)
\bibitem{co} A. Connes,  Non commutative geometry. Academic
Press, Inc. London (1990)
\bibitem{la} G. Landi,  An introduction to noncommutative
spaces and their geometries. Springer-Verlag (1997).
\bibitem{mssw}
J. Madore, S. Schraml, P. Schupp and J. Wess,  Eur.  Phys. J.  C16, 161 (2000)
\bibitem{bcz}
D. Brace, B. L. Cerchiai and B. Zumino,  Int. J. Mod. Phys. A
1751, 205 (2002)
\bibitem{av}
L. Alvarez-Gaume and M. A. Vazquez-Mozo, Nucl. Phys. B668, 293 (2003)
\bibitem{ov} H. Ooguri and C. Vafa, 	Adv. Theor. Math. Phys. 7, 53 (2003)
\bibitem{se} N. Seiberg, JHEP.  0306, 010 (2003)
\bibitem{bgn} J. de Boer, P. A. Grassi and P. van Nieuwenhuizen,
 Phys. Lett. B574, (2003) 98 (2003)
\bibitem{beta1}
E. Chang-Young, H.Kim and H. Nakajima, JHEP. 0804, 004 (2008) 
\bibitem{beta2}
K. Araki, T.  Inami, H. Nakajima and Y. Saito, JHEP. 0601, 109 (2006) 
\bibitem{beta3}
J. S. Cook, J. Math. Phys. 47, 012304  (2006)
\bibitem{beta4}
Y. Kobayashi and S. Sasaki, Phys. Rev. D72, 065015  (2005) 
\bibitem{bs} N. Berkovits and N. Seiberg, JHEP. 0307, 010 (2003)
\bibitem{14om}
C. S. Chu and E. Sezgin,  JHEP. 9712, 001
(1997) 
\bibitem{abjm1234}
A. F. Ferrari, M. Gomes, J. R. Nascimento, A. Yu. Petrov and A. J. da Silva,
Phys. Rev. D74,  125016, (2006)
\bibitem{41s}
P. S. Howe and E. Sezgin,  Phys. Lett. B 394, 62 (1997) 
\bibitem{41ss}
M. Aganagic, J. Park, C. Popescu and J. H. Schwarz,  Nucl. Phys. B 496, 191 (1997)
\bibitem{41ssa}
P. Pasti, D. P. Sorokin and M. Tonin,  Phys. Lett. B 398, 41 (1997) 
\bibitem{41ssb}
I. A. Bandos, K. Lechner, A. Nurmagambetov, P. Pasti, D. P. Sorokin and M. Tonin,
 Phys. Rev. Lett. 78, 4332 (1997)
\bibitem{42s}
P. S. Howe, E. Sezgin and P. C. West, Phys. Lett. B 399, 49 (1997)
\bibitem{e}
M. Fujita, W. Li, S. Ryu and T. Takayanagi,  JHEP. 0906, 066 (2009)
\bibitem{6}I. R. Klebanov and A. M. Polyakov, Phys. Lett. B550, 213 (2002)
\bibitem{7}J. H. Schwarz, JHEP. 0411, 078 (2004)
 \bibitem{8}C. Ahn, H. Kim, B. H. Lee and H. S. Yang, Phys. Rev. D61, 066002
(2000)
 \bibitem{9}B. Chen and J. B. Wu, JHEP. 096, 0809 ( 2008)
 \bibitem{10}M. Benna, I. Klebanov, T. Klose and M. Smedback, JHEP. 0809, 072
(2008)
\end{thebibliography}
\end{document}